\documentclass[letterpaper,twocolumn,showpacs,
prb,floatfix
]{revtex4}

\usepackage{graphicx}
\usepackage{amssymb}
\usepackage{amsmath,amsfonts,latexsym}
\usepackage{array,tabularx,color}
\usepackage[normalem]{ulem}

\DeclareMathOperator{\tr}{tr}
\newcommand{\vect}[1]{\mathbf{#1}}
\newcommand*{\cL}{{\cal L}}
\newcommand*{\br}{\mathbf{r}}
\newcommand*{\bp}{\mathbf{p}}
\newcommand*{\bq}{\mathbf{q}}

\begin{document}

\title{Spinodal decomposition in polarised Fermi superfluids}

\author{A. Lamacraft} 
\email{austen@virginia.edu}
\affiliation{Rudolf Peierls Centre for Theoretical Physics, 1 Keble
Road, Oxford OX1 3NP, UK}

\author{F. M. Marchetti} 
\email{fmm25@cam.ac.uk} 
\affiliation{Rudolf Peierls Centre for Theoretical Physics, 1 Keble
Road, Oxford OX1 3NP, UK}
 
\date{January 28, 2007}       

\begin{abstract}
  We discuss the dynamics of phase separation through the process of
  spinodal decomposition in a Fermi superfluid with population
  imbalance. We discuss this instability first in terms of a
  phenomenological Landau theory. Working within the mean-field
  description at zero temperature, we then find the spinodal region in
  the phase diagram of polarisation versus interaction strength, and
  the spectrum of unstable modes in this region. After a quench, the
  spinodal decomposition starts from the Sarma state, which is a
  minimum of the free energy with respect to the order parameter
  \emph{at fixed density and polarisation} and a maximum at fixed
  chemical potentials. The possibility of observing non-trivial domain
  structures in current experiments with trapped atomic gases is
  discussed.
\end{abstract}

\pacs{03.75.Kk, 03.75.Ss, 64.75.+g}

\maketitle

The ordering of matter into different phases is a central
preoccupation of many areas of physics, from condensed matter to
cosmology. Hand in hand with the \emph{existence} of different phases
goes the question of the dynamical processes responsible for their
formation, which may be equally important in determining what is
observed in a given situation. Recent experimental advances in the
creation of degenerate atomic gases have begun to realize the
prospect of a rich variety of new phases in atomic matter, involving
the hyperfine degrees of freedom, mixtures of different species, or
spatial order on optical lattices. With each new phase comes the
dynamical issue of how that phase will appear under laboratory
conditions. One advantage offered by atomic systems is that the
characteristic timescale $\hbar/k_BT$ at nanokelvin temperatures is in
the convenient millisecond range.

The possibility of tuning interparticle interactions in a controlled
manner has proven to be of particular significance
lately. Magnetically tuned Feshbach resonances have permitted the
experimental investigation of the crossover from a Bose-Einstein
condensate (BEC) of diatomic molecules to the
Bardeen-Cooper-Schrieffer (BCS) limit of weakly-bound Cooper pairs of
fermionic
atoms~\cite{regal2004,zwierlein2004,chin2004,bourdel2004,kinast2004,zwierlein2005}.

It appears that when the numbers of atoms of the two species
undergoing pairing are equal, the system forms a condensate with
smoothly varying properties at low temperatures. With unequal numbers
(we will call such a system `polarised') there is the possibility of
phase separation into a superfluid of low polarisation (favoured by
pairing) and a normal fluid of higher
polarisation~\cite{zwierlein2006,partridge2006,zwierlein2006_2,shin2006,partridge2006_2,bedaque2003,sheehy2006}.

\begin{figure} 
\centering \includegraphics[width=0.46\textwidth]{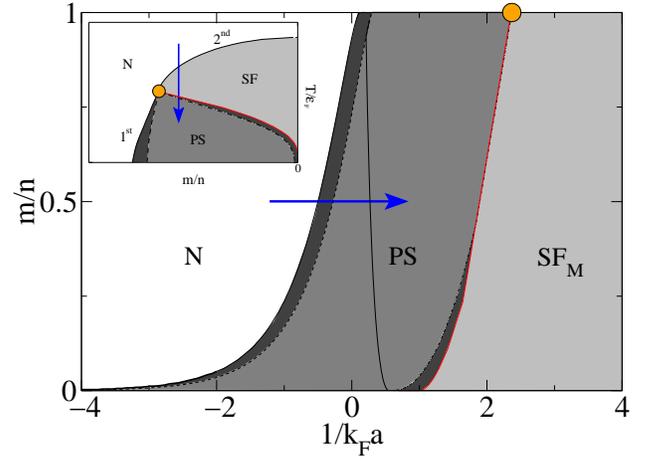}
\caption{(Colour online) Zero temperature mean-field phase diagram for
         magnetisation $m/n$ versus interaction strength $1/k_Fa$. The
         phase separated region (PS) can be decomposed into two
         regions --- the spinodal or unstable region and the
         metastable region (darker shade) --- divided by the spinodal
         normal (sp-N) line (dashed) and the spinodal superfluid
         (sp-SF) line (dot-dashed). In addition, the PS spinodal phase
         is divided in half by the line where the superfluid density
         $Q$ is zero (thin solid). The tricritical point (orange
         circle) is at $m/n=1$ and $1/k_Fa\simeq 2.37$. Inset:
         schematic finite temperature phase diagram of
         $T_c/\varepsilon_F$ versus $m/n$ for a fixed interaction
         strength $1/k_Fa$. Arrows indicate possible quenches into the
         spinodal region.}
\label{fig:spinodals}
\end{figure}

With the occurrence of phase separation in these systems now
established, it is crucial to examine the dynamics of this process,
and that is the purpose of this Letter. Working within a mean-field
approximation, we find that the dome of phase separation at
temperatures below the tricritical point~\cite{parish2006,gubbels2006}
contains a \emph{spinodal} region where phase separation proceeds via
a linear instability. In many recent papers this region is incorrectly
identified with the phase coexistence
region~\cite{chien2006,iskin2006,pao2006,pao2006b} (see
Fig.~\ref{fig:spinodals}).
%
%

We discuss the dynamics initiated by a quench into the spinodal region
by finding the unstable density modes associated with the endpoint of
the quench. The modes with the fastest growth rate give the
characteristic size of the resulting domains of superfluid and normal
fluid. Though our description of these processes will not be
quantitatively correct in the crossover region where the system is
strongly interacting, we emphasise that this behaviour is a generic
feature of systems possessing this type of phase diagram. In
particular, our system shares many similarities with the problem of
$^{3}$He-$^{4}$He mixtures (see e.g. \cite{gunton,hoffer1986}). We
will confine ourselves to the early stages of spinodal decomposition
characterized by the exponential growth of unstable modes. The
emergence of a coarsening regime at later times, where domains scale
with the time since the quench~\cite{bray1994}, is a fascinating
possibility that we will leave for future work.

The rest of this paper is organized as follows. In the next section we introduce a simple phenomenological model for the polarized system. We use this model to show how, with a simplifying assumption, mean-field calculations within the grand canonical ensemble may be applied to the early stages of spinodal decomposition. By adding dynamical assumptions, the equation satisfied by the sound velocity can be inferred. In section~\ref{sec:micro} we provide a microscopic mean-field calculation of the spectra of unstable modes, which are discussed in some detail in section~\ref{sec:discuss} before we conclude.

\section{Phenomenological model}

We begin by discussing the phase
diagram in phenomenological terms starting from a model free energy
depending on the superfluid order parameter $\Delta$ and the density
difference $m=n_{\uparrow}-n_{\downarrow}$, following a similar approach to the $^{3}$He-$^{4}$He system~\cite{hohenberg1979}
\begin{multline}
  f_m (\Delta,m) \equiv \frac{r}{2} |\Delta|^2 + u |\Delta|^4 +
  v|\Delta|^6 \\ + \frac{1}{2}\chi_n^{-1} m^2+\gamma m|\Delta|^2 \; .
\label{landau_model}
\end{multline}
The potential for $\Delta$ is the simplest one that can describe a
first order transition. We are interested in $r<0$ so that for $m=0$
there is always superfluid order $\Delta\neq 0$. Further, the coupling
to $m$ has the obvious physical meaning that larger polarisations
discourage pairing. We ignore for the moment additional couplings to
the total density $n=n_{\uparrow}+n_{\downarrow}$~\footnote{One can in
general show that, at the mean-field level, the free
energy~\eqref{landau_model} is a function of $m/n$ only, and
conversely the grand canonical potential~\eqref{landau_delta} is
function of $h/|\mu|$ only.}.

Minimising~\eqref{landau_model} on the order parameter, gives the
potential
\begin{equation}
  F(m)\equiv \min_\Delta f_m(\Delta,m) \; . 
\label{eq:mpot}
\end{equation}
Where the transition is first order, the phase coexistence region
$(m_{\text{cr-SF}},m_{\text{cr-N}})$ is obtained by the usual tangent
construction~\cite{callen}. Inside the coexistence region one
identifies a spinodal region $(m_{\text{sp-N}},m_{\text{sp-SF}})$,
where the susceptibility $\partial_m^2 F(m)<0$, and phase separation
proceeds via the growth of unstable modes. In general this represents
an extremely difficult dynamical problem. We will make the simplifying
assumption that following a quench into the spinodal region, the order
parameter relaxes rapidly do its minimum at fixed $m$, while
$m(\vect{r})$, being a conserved quantity, begins to develop
inhomogeneities on a much longer timescale. For the case of trapped
gases, we will review the validity of this approach \emph{a posteriori}.

\begin{figure} 
\centering
\includegraphics[width=0.46\textwidth]{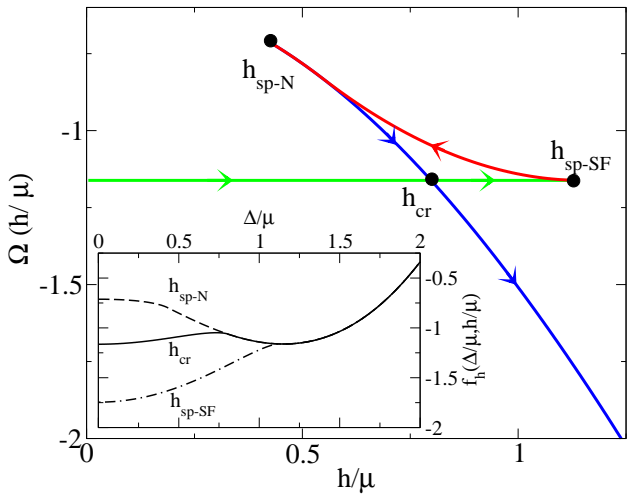}
\caption{(Colour online) Mean-field free energy $\Omega (h/\mu)$ for
         the Hamiltonian~\eqref{eq:ham} versus $h/\mu$ at unitarity
         $1/k_Fa=0$ ($\mu>0$) obtained evaluating the free energy
         density $f_h (\Delta/\mu,h/\mu)$ respectively at the local
         minimum for $\Delta/\mu\simeq 1.15$ (green), at $\Delta=0$
         (blue) and at the local maxima (red), i.e. the Sarma
         state. The value of $h/\mu$ corresponding to the
         spinodal-normal point (sp-N), the spinodal-superfluid point
         (sp-SF) and the critical value for phase separation (cr) are
         indicated. At unitarity, the corresponding values of the
         magnetisation are respectively $m_{\text{cr-SF}}/n =
         m_{\text{sp-SF}}/n = 0$, $m_{\text{sp-N}}/n \simeq 0.73$ and
         $m_{\text{cr-N}}/n \simeq 0.93$ (see
         Fig.~\ref{fig:spinodals}). Inset: Plots of the free energy
         density $f_h (\Delta/\mu,h/\mu)$ for the values of $h$
         corresponding to $h_{\text{sp-N}}$, $h_{\text{sp-SF}}$ and
         $h_{\text{cr}}$.  Note that the thermodynamic free energy
         corresponding to the Sarma state has a positive curvature
         indicating instability. The cusp structure shrinks to a point
         at the tricritical point.}
\label{fig:spinodal}
\end{figure}

Often it is convenient, particularly in many body calculations, to
work instead with the grand canonical potential $f_h(\Delta,h)$,
obtained from $f_m$ in the usual way
\begin{multline}
  f_h(\Delta,h) \equiv \min_m \left[f_m(\Delta,m) - hm\right] \\
  =\frac{1}{2}\tilde{r} |\Delta|^2+ \tilde{u}|\Delta|^4 +
  v|\Delta|^6-\frac{1}{2}\chi_nh^2 \; ,
\label{landau_delta}
\end{multline}
where $\tilde{r}= r+2\gamma h\chi_n$ and $\tilde{u} =
u-\frac{1}{2}\chi_n \gamma^2<0$. If we assume $u>0$, then for small
$\gamma$ a second order transition occurs when $\tilde r$ changes
sign. Increasing $\gamma$ causes the transition to become first order
as $\tilde u$ becomes negative. Minimising $f_h(\Delta,h)$ on $\Delta$
gives the thermodynamic free energy $\Omega(h) \equiv \min_\Delta
f_h(\Delta,h)$. At some critical $h=h_{\text{cr}}$ there is a
discontinuity in the derivative (see Fig.~\ref{fig:spinodal}). In a
uniform phase $m=- \partial_h \Omega$ so the boundaries of the phase
coexistence region are $m_{\text{cr-N,cr-SF}}=-\partial_h
\Omega|_{h_\text{cr}^{\pm}}$. The two phases correspond to the two
minima of $f_h(\Delta,h)$: one at $\Delta=0$, the normal phase, and
the superfluid phase at finite $\Delta$. One can continue past
$h_{\text{cr}}$ on the metastable minimum, rather than the true
minimum. Such states are linearly stable, with a positive
susceptibility $- \partial_h^2\Omega$ ($= [\partial_m^2 F(m)]^{-1}$),
since one may easily see that
\begin{equation}
  \frac{\partial^2\Omega}{\partial h^2} =
    \frac{\partial^2f_h}{\partial h^2} -
    \left(\frac{\partial^2f_h}{\partial \Delta^2}\right)^{-1}
    \left(\frac{\partial m}{\partial \Delta}\right)^2\; ,
\label{sus_rel}
\end{equation}
and the first term on the right hand side is $-\chi_n<0$ --- the
normal state susceptibility is assumed positive. Since the curvature
of $\Omega(h)$ remains negative, $m$ takes values inside the
coexistence region but the system remains uniform. This is of course
not the equilibrium state of the system, but the other phase must
nucleate in order for phase separation to occur.

%
%

This situation changes when the metastable minima merge with the
maximum, for $h$ corresponding respectively to $m_{\text{sp-N}}$ and
$m_{\text{sp-SF}}$ for the metastable superfluid and normal phases. As
already discussed, inside the spinodal region we seek the
\emph{constrained minimum} of $f_m(\Delta,m)$ with respect to $\Delta$
at fixed $m$, and the spectrum of unstable modes about this point. How
should this programme be implemented for a many-body calculation that
provides instead the grand canonical potential $f_h(\Delta,h)$?
Fortunately, one may easily show that the stationary points of $f_m
(\Delta,m)$ with respect to $\Delta$ coincide with those of
$f_h(\Delta,h)$ \emph{at corresponding values of $h$ and $m$}. This is
because $f_m(\Delta,m)$ may be written as
\begin{equation*}
  f_m(\Delta,m)=f_h(\Delta,h^*(\Delta,m))+h^*(\Delta,m)m \; ,
\label{Fform}
\end{equation*}
where $h^*(\Delta,m)$ is the value of $h$ that maximises
$f_h(\Delta,h)+hm$. Thus $m=-\partial_h\Omega|_{h=h^*}$. We can then
easily see that the conditions, $\partial_\Delta f_m(\Delta,m) = 0$
and $\partial_\Delta f_h(\Delta,h)=0$ are equivalent when
$h=h^*(\Delta,m)$.

In the spinodal region the unstable constrained minimum corresponds to
a \emph{maximum} of $f_h$, as may be seen from Eq.~\eqref{sus_rel}:
$\partial^2\Omega/\partial h^2>0$ requires $\partial^2f_h/\partial
\Delta^2<0$. In the context of the mean-field theory for the paired
fermion system, this corresponds to the solution of the
self-consistent equation in a magnetic field discovered by
Sarma~\cite{sarma1963}.

These results are readily adapted to the presence of a trap potential
$V(\vect{r})$ using the following simple approximation: At each point
in space we find the \emph{constrained minimum} of the order parameter
for fixed local magnetisation $m(\vect{r})/n(\vect{r})$ corresponding
to the density profiles before the quench. Spinodal decomposition
therefore occurs where the local magnetisation lies in the spinodal
region of the homogeneous phase diagram. Moving out from the centre of
the trap corresponds to a vertical trajectory in
Fig.~\ref{fig:spinodals} which will be displaced horizontally by a
quench in $1/k_Fa$.

To make contact with the microscopic analysis of the next section, let us add appropriate dynamics to the above model. Expanding $\Delta_0+\delta\Delta$ around some value we write a phenomenological quadratic Lagrangian for the longitudinal (amplitude) and transverse (phase) modes
\begin{eqnarray}\label{model_L}
\cL=B\delta\Delta_L\dot{\delta\Delta_T}+\frac{R}{2}(\dot{\delta\Delta_T})^2-\frac{Q}{2}\left(\nabla\delta\Delta_T\right)^2\nonumber\\-\frac{A}{2}(\delta\Delta_L)^2-2\gamma m\Delta_0\delta\Delta_L.
\end{eqnarray}
Eq.~(\ref{model_L}) includes all terms up to second order in the derivatives except for a $(\dot{\delta\Delta_T})^2$ term that will not change the sound velocity. $\delta\Delta_L$ is coupled to the density difference $m$ as specified in the model Eq.~(\ref{landau_model}). $m$ is written in terms of the distribution function of the majority quasiparticle distribution function (at zero temperature there are no minority quasiparticles)
\[m(\br)=\sum_\bp n_\uparrow(\bp,\br),\]
which obeys the Boltzmann equation
\[
\left[\partial_t +\vect{v}_F\cdot\nabla_{\br}-2\gamma \Delta_0\nabla_\br\delta\Delta_L\cdot \nabla_\bp\right]n_{\uparrow}(\bp,\br,t)=0.
\]
where $\vect{v}_F=v_F\hat\bp$ is the Fermi velocity. The linearized solution is expressed as
\begin{eqnarray*}
n_{\uparrow}(\bp,\bq,\omega)=\frac{\bq\cdot\vect{v}_F}{w-\vect{v}_F\cdot\bq}2\gamma \Delta_0\delta\Delta_L(\bq,\omega)\delta(\epsilon_\bp-\mu)\nonumber\\
m(\bq,\omega)=2\gamma \Delta_0\nu(\mu)L(\omega/|\bq|)\delta\Delta_L(\bq,\omega)
\end{eqnarray*}
where $L(x)=\frac{x}{2}\log\frac{x+1}{x-1}-1$ is the Lindhard function, and $\nu(\mu)$ the Fermi surface density of states. The dispersion relation of the linearized modes is then given by a solution of 
\begin{eqnarray*}
\left|\begin{array}{ccc}A & iB\omega & 2\gamma\Delta_0 \\-iB\omega & Qq^2-R\omega^2 & 0 \\2\gamma\Delta_0\nu(\mu)L(\omega/|\bq|) & 0 & -1\end{array}\right|=0, 
\end{eqnarray*}
or 
\begin{equation}\label{model_dispersion}
\left(A+4\gamma^2\Delta_0^2\nu L(c_s)\right)\left(Q-Rc_s^2\right)-B^2c_s^2=0.
\end{equation}
We will see that an equation of the same form emerges from the microscopic analysis of the next section, where the solutions will be further analyzed.

\section{Microscopic calculation}\label{sec:micro}

We turn now to the analysis of the microscopic problem described by
the Hamiltonian
\begin{multline}
  \hat{H} - \sum_{\sigma = \uparrow,\downarrow} \mu_\sigma
  \hat{n}_{\sigma}  = \sum_{\vect{k}\sigma}
  \left(\epsilon_{\vect{k}} - \mu_{\sigma}\right)
  c_{\vect{k} \sigma}^\dag c_{\vect{k} \sigma}^{\vphantom{\dag}} \\
   + \frac{g}{V} \sum_{\vect{k},\vect{k}',\vect{q}}
  c_{\vect{k}+\vect{q}/2 \uparrow}^\dag c_{-\vect{k}+\vect{q}/2
  \downarrow}^\dag c_{-\vect{k}'+\vect{q}/2 \downarrow}^{\vphantom{\dag}}
  c_{\vect{k}'+\vect{q}/2 \uparrow}^{\vphantom{\dag}}\; ,
\label{eq:ham}
\end{multline}
where $\epsilon_{\vect{k}}=k^2/2m$ (we set $\hbar=1$) and where the
scattering length is introduced in the usual way:
\begin{equation}
  \frac{1}{g} = \frac{m}{4\pi a} - \frac{1}{V} \sum_{\vect{k}}
  \frac{1}{2\epsilon_{\vect{k}}} \; .
\end{equation}
The condition $\partial^2f_h/\partial \Delta^2=0$, corresponding to a
divergent susceptibility, was used to obtain the spinodal lines in
Fig.~\ref{fig:spinodals}. As explained before, the unstable modes are
to be found from the matrix response function (dynamical
susceptibility):
\begin{multline}
  \hat\chi(\vect{r}-\vect{r}',t-t') \\ = -i\begin{pmatrix}
  \langle\left[n(\vect{r},t),n(\vect{r}',t')\right]\rangle &
  \langle\left[n(\vect{r},t),m(\vect{r}',t')\right]
  \rangle\\\langle\left[m(\vect{r},t),n(\vect{r}',t')\right]\rangle &
  \langle\left[n(\vect{r},t),m(\vect{r}',t')\right]\rangle\end{pmatrix}\; .
\label{eq:response}
\end{multline}
Finding the spectrum of collective modes requires us to solve the
equation
\begin{equation}
  \det \hat{\chi}^{-1}(\vect{q},\varepsilon (\vect{q})) = 0,
\label{det}
\end{equation}
which defines the dispersion relation $\varepsilon(\vect{q})$. The
unstable modes correspond to $\Im \varepsilon(\vect{q}) > 0$. In
practice, the response matrix~\eqref{eq:response} can be found making
use of a path integral formulation, by expanding the action
\begin{gather}
  S[\Delta,\mu,h] = -\frac{1}{g} \int_0^{\beta} d\tau \int d\vect{r}
  |\Delta|^2 - \tr \ln \hat{\mathcal{G}}^{-1}\\\nonumber
  \hat{\mathcal{G}}^{-1} =
  \begin{pmatrix} \partial_{\tau} -\frac{\nabla^2}{2m} -\mu - h & -\Delta
  \\  -\Delta^* & \partial_{\tau} +\frac{\nabla^2}{2m} +\mu -
  h\end{pmatrix} \; ,
\end{gather}
up to second order in fluctuations of $\mu (\vect{r},\tau) = \mu +
\delta \mu (\vect{r},\tau)$, $h (\vect{r},\tau) = h + \delta h
(\vect{r},\tau)$ and $\Delta(\vect{r},\tau) = \Delta + \delta
\Delta^{L} (\vect{r},\tau) + i \delta \Delta^{T} (\vect{r},\tau)$
around their mean-field values. Here, we have introduced $\mu =
(\mu_{\uparrow} + \mu_{\downarrow})/2$ and $h = \mu_{\uparrow} -
\mu_{\downarrow}$.

By completing the squares in
$\delta\Delta^{L}$ and $\delta \Delta^{T}$, one can easily obtain:
\begin{multline}
  \hat{\chi} (\vect{q},i\omega_h) =\begin{pmatrix} \Pi_{\mu\mu} &
  \Pi_{h \mu} \\\Pi_{\mu h} & \Pi_{h h}\end{pmatrix}\\
  -\begin{pmatrix}\Pi_{\mu\Delta^L} & \Pi_{\mu \Delta^T} \\\Pi_{h
  \Delta^L} & \Pi_{h \Delta^T}\end{pmatrix} \hat{\mathcal{D}}
  \begin{pmatrix}\Pi_{\Delta^L\mu} & \Pi_{\Delta^L h}
  \\\Pi_{\Delta^T \mu} & \Pi_{\Delta^T h}\end{pmatrix}\; ,
\label{mf_resp}
\end{multline}
where we have introduced the polarization operators
\begin{equation*}
  \Pi_{ab} (\vect{q},i\omega_h) = \frac{1}{V\beta}
  \sum_{\vect{k},i\epsilon_n} \tr \,\tau_a
  \hat{G}_{\vect{q}+\vect{k},i\epsilon_n+i\omega_h}^{(0)}\tau_b
  \hat{G}_{\vect{q},i\epsilon_n}^{(0)} \; ,
\end{equation*}
with $\tau_\mu=\sigma_3$, $\tau_h=\openone$,
$\tau_{\Delta^L}=\sigma_1$, and $\tau_{\Delta^T}=\sigma_2$, and where
the order parameter propagator is
\begin{equation}
  \hat{\mathcal{D}}^{-1} = -\frac{2}{g} \openone + \begin{pmatrix}
  \Pi_{\Delta^L\Delta^L} & \Pi_{\Delta^L \Delta^T} \\\Pi_{\Delta^T
  \Delta^L} & \Pi_{\Delta^T \Delta^T}\end{pmatrix} \; .
\end{equation}
We omit the explicit expressions for these quantities, as they are
straightforward generalisations of the expressions found e.g. in
Ref.~\onlinecite{engelbrecht1997} to the case $h\neq 0$. It is clear that
Eq.~\eqref{mf_resp} is the generalisation of Eq.~\eqref{sus_rel} to
the full matrix response and to nonzero $\vect{q}$ and $\varepsilon$.

%
\begin{figure} 
\centering
\includegraphics[width=0.46\textwidth]{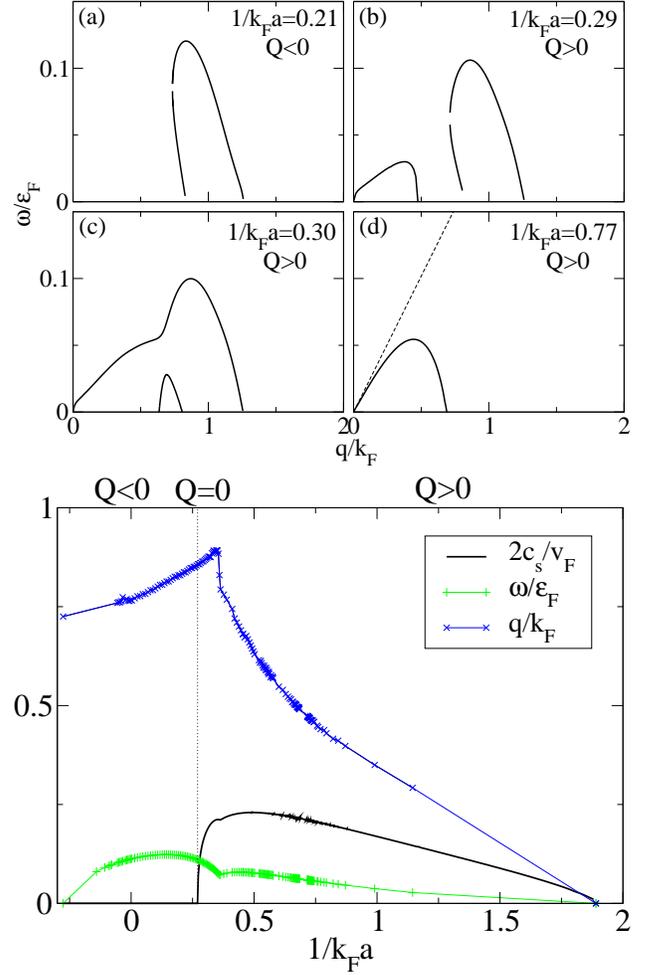}
\caption{(Colour online) Upper panels: Unstable modes frequency
          $\omega/\varepsilon_F$ ($\omega(\vect{q}) \equiv
          -i\varepsilon(\vect{q})$) versus momentum $q/q_F$ for
          different values of the interaction strength $1/k_Fa$ across
          the spinodal region and for fixed polarisation
          $m/n=0.5$. Note that gaps correspond to complex
          frequencies. Lower panel: Plot of the most unstable mode
          frequency $\omega/\varepsilon_F$ (plus green), momentum
          $q/q_F$ (times blue) and of the sound velocity $2c_s/v_F$
          (solid black) across the spinodal region for fixed
          polarisation $m/n=0.5$.}
\label{fig:unstable}
\end{figure}
%

The mean-field approximation to the mode spectrum is the solution of
Eq.~\eqref{det}, with the response matrix given by the
expression~\eqref{mf_resp}. Here, $\mu$ and $h$ are chosen so that $n$
and $m$ take the desired values, and inside the spinodal region,
$\Delta$ is taken at the Sarma value corresponding to the maximum of
$f_{h} (\Delta,h)$ --- as discussed, the stationary points of $f_{h}
(\Delta,h)$ coincide with those of $f_m (\Delta, m)$. Evidently, a
sufficient condition for a solution of Eq.~\eqref{det} is the
occurrence of a pole at $\varepsilon(\vect{q})$ in the order parameter
propagator, so we must solve $\det
\hat{\mathcal{D}}^{-1}(\vect{q},\varepsilon(\vect{q}))=0$. In general,
a numerical solution is called for. At the spinodal lines, however, a
diverging susceptibility implies a vanishing sound velocity, which we
can find by expanding in $\varepsilon$ and $\vect{q}$:
\begin{equation}
  \hat{\mathcal{D}}^{-1} \simeq \begin{pmatrix} A+P(\varepsilon/q) &
  iB\varepsilon \\-iB\varepsilon & Qq^2-R\varepsilon^2 \end{pmatrix}
  \; ,
\label{eq:expan}
\end{equation}
where $A= \frac{1}{V} \sum_{\vect{k}}
  \Theta(E_\vect{k}-h)\frac{\Delta^2}{2E_\vect{k}^3}$, $ B =
  \frac{1}{V} \sum_{\vect{k}}
  \Theta(E_\vect{k}-h)\frac{\epsilon_{\vect{k}} -\mu}{4E_\vect{k}^3}$,
  $R=\frac{1}{V} \sum_{\vect{k}}
  \frac{\Theta(E_\vect{k}-h)}{8E_\vect{k}^3}$ and 
\begin{gather*}  
  Q = \frac{n}{8m\Delta^2}\left[1-\frac{h}{2
  \sqrt{h^2-\Delta^2}}\frac{\epsilon_+^{3/2}  +\epsilon_-^{3/2}}{\epsilon_F^{3/2}}\right]\\
  P(\varepsilon/q) = \frac{\Delta^2}{2h^2} \left[\nu_{+}
  L(\varepsilon/v_{+} q) + \nu_{-} L(\varepsilon/v_{-} q)\right] \; ,
\end{gather*}
with $E_{\vect{k}} = \sqrt{(\epsilon_{\vect{k}} -\mu)^2+\Delta^2}$,
 and where $\nu_{\pm}=\frac{m \epsilon_{\pm}}{\pi^2}v_{\pm}^{-1}$ and
$v_{\pm}=dE_k/dk|_{\epsilon_{\pm}}$ are the density of states and
velocity at the two solutions $\epsilon_{\pm}$ of $E_{\vect{k}}=h$
(take $\epsilon_{\pm}=0=\nu_{\pm}$ if there is no solution). Note that the phase
stiffness $4 m \Delta^2 Q$ is the superfluid density~\cite{pao2006},
which changes sign inside the spinodal region (see
Fig.~\ref{fig:spinodals}).

The higher order terms in the $(1,1)$ entry of
Eq.~\eqref{eq:expan} do not affect the sound velocity $c_s$, which is
the solution of
\begin{equation}\label{true_sound}
  \left[A + P(c_s)\right] \left(Q - Rc_s^2\right) - B^2c_s^2 =0 \; .
\end{equation}
For obvious reasons, this analysis closely parallels that of
Ref.~\onlinecite{santamore2004} for Bose-Fermi mixtures, and Eq.~\ref{true_sound} reproduces the form of Eq.~\ref{model_dispersion} found earlier.

\section{Discussion}\label{sec:discuss}

At the spinodal lines $A+P(0)=0$. This is the same condition that was
used in Refs.~\onlinecite{chien2006,chen2006} to identify the unstable
region, although the possibility of metastability was
ignored. $A+P(0)<0$ inside the spinodal region, and one can
distinguish two cases: When $Q>0$ the sound velocity $c_s$ is pure
imaginary, while for $Q<0$ it is in general complex. As a consequence,
the unstable mode spectrum changes in character as one moves across
$Q=0$ (see Fig.~\ref{fig:unstable}). Approaching the region of negative superfluid density $Q<0$
from the region $Q>0$, a second imaginary modes appears (see panel (c)
in Fig.~\ref{fig:unstable}) and the two merge (panel (b)): The
resulting `gap' region corresponds to complex frequencies, implying a
`flickering' component to the instability. The most
unstable mode always corresponds to a pure imaginary
frequency, however.
While the most unstable mode frequency and wavevector go to zero on
the superfluid side of the spinodal region, the instability towards a
Fulde-Ferrell-Larkin-Ovchinnikov (FFLO) phase means that the characteristic
wavevector does not go to zero on the normal
side~\cite{sheehy2006}. Indeed, one may view the early stages of
spinodal decomposition as a transient FFLO state.

The characteristic length and time scales at which inhomogeneities
appear as the precursor of phase separation are determined by the most
unstable modes. At unitarity $1/k_Fa=0$ and for $T_F=1\mu$K, this time
scale is of order of $400\mu$s, which is on the same scale as the
condensate formation time, as measured in
Ref.~\onlinecite{zwierlein-form}. The length scale is roughly $1/k_F \simeq
0.1\mu$m. Both scales become larger as one approaches the spinodal
lines, which will always occur somewhere in the trap. The unitarity
region may be a suitable place to observe the late stages of spinodal
decomposition and the possible existence of a coarsening regime.

In conclusion, we have studied the early stage dynamics of phase
separation in polarised Fermi superfluids. We expect that the
investigation of these instabilities is within reach of current
experiments.

We are grateful to P. Eastham for help with the numerics and to TCM
  group in Cambridge for the use of computer resources. FMM would like
  to acknowledge the financial support of EPSRC.


 \end{document}